\documentclass[12pt,preprint]{aastex}
\newcommand {\hii}{H\,{\sc ii}} 
\newcommand {\kms}{\relax \ifmmode {\,\rm km\,s}^{-1}\else \,km\,s$^{-1}$\fi}

\newcommand {\hd}{HD\,5980}
\newcommand {\n}{[NHS03]}
\newcommand {\x}{{\it XMM-Newton}}
\newcommand {\ro}{{\it ROSAT}}
\newcommand {\ch}{{\it Chandra}}

\usepackage{rotating}

\shorttitle{NGC\,346}
\shortauthors{Naz\'e et al.}

\begin{document}

\title{An X-ray investigation of the NGC\,346 field in the SMC (3): \x\ data}

\author{Ya\"el Naz\'e\altaffilmark{1} and Jean Manfroid\altaffilmark{2,3}}
\affil{Institut d'Astrophysique et de G\'eophysique,
Universit\'e de Li\`ege, All\'ee du 6 Ao\^ut 17, Bat. B5c, B4000 -
Li\`ege (Belgium)}
\email{naze@astro.ulg.ac.be, manfroid@astro.ulg.ac.be} 

\author{Ian R. Stevens}
\affil{School of Physics \& Astronomy, University of
Birmingham, Edgbaston, Birmingham B15 2TT (UK)}
\email{irs@star.sr.bham.ac.uk} 

\author{Michael F. Corcoran}
\affil{Universities Space Research Association, 7501 Forbes
Blvd, Ste 206, Seabrook, MD 20706, and Laboratory for High Energy 
Astrophysics, Goddard Space Flight Center, Greenbelt, MD 20771 (USA)}
\email{corcoran@barnegat.gsfc.nasa.gov} 

\author{Aaron Flores}
\affil{Universidad Autonoma del Carmen, Facultad de Ingenieria, Calle 56 No 4, Cd del Carmen, Campeche (Mexico)}
\email{aflores@pampano.unacar.mx}

\altaffiltext{1}{Research Fellow F.N.R.S.}
\altaffiltext{2}{Research Director F.N.R.S.}
\altaffiltext{3}{Visiting Astronomer, European Southern Observatory}

\begin{abstract}

We present new \x\ results on the field around the NGC\,346 star cluster
in the SMC. This continues and extends previously published work on {\it
Chandra} observations of the same field. The two \x\ observations were
obtained, respectively, six months before and six months after the
previously published  {\it Chandra} data.
Of the 51 X-ray sources detected with \x, 29 were already detected with
{\it Chandra}. Comparing the properties of these X-ray sources in each
of our three datasets has enabled us to investigate their variability on
times scales of a year. Changes in the flux levels and/or spectral
properties were observed for 21 of these sources. In addition, we
discovered long-term variations in the X-ray properties of the peculiar
system \hd, a luminous blue variable star, that is likely to be a
colliding wind binary system, which displayed the largest luminosity
during the first \x\ observation.


\end{abstract}


\section{Introduction}

The giant \hii\ region N66 of the Small Magellanic Cloud (SMC)
is the largest star formation region of that galaxy. It notably
harbors NGC\,346, a young cluster containing a wealth of massive stars,
and \hd, a peculiar system that underwent a LBV-type eruption at
the end of the last century. \ro\ and {\it ASCA} observations
have also revealed the presence of a few X-ray binaries (XRBs)
in this field \citep[see e.g. ][]{ha00,ts99}, some of which were later
found to harbor pulsars \citep[for a review, see ][]{ha04}. 

Recently, a new generation of powerful X-ray observatories has been 
launched and the 
XMEGA\footnote{http://lheawww.gsfc.nasa.gov/users/corcoran/xmega/xmega.html} 
consortium used this opportunity to observe the X-ray emission from the 
NGC\,346 field in greater detail.
The results of a 100~ks $Chandra$ observation have been
given in two previous articles. In Naz\'e et al. (2003a, hereafter
Paper I), we reported the first detections in the X-ray domain
of the cluster and \hd. The X-ray emission from NGC\,346 appeared 
tightly correlated with the cluster's core, while that of \hd\ was 
found to be bright and variable on the short timescale.
In Naz\'e et al. (2003b, hereafter Paper II), we analyzed 
the X-ray properties of the 75 point sources discovered in the 
field, and found possible counterparts to 32 of these sources. 
We refer the reader to these papers for detailed results and
a thorough introduction to the importance of the NGC\,346 field. 
This third Paper of the series is meant to
supplement the two previous ones. Here, we continue our 
investigation of the field with the analysis of two \x\ datasets,
which provide a wider field compared to the {\it Chandra} data and allow
us to examine the issue of source variability on timescales of seconds to 
months.

The observations and data reduction are presented in \S2. Next, the 
properties of the sources detected by \x\ are discussed in \S3, where
they are also compared to previous data available. In \S4, we focus 
the discussion on \hd, the supernova remnant SNR\,0056$-$72.5, and the 
XRBs. We conclude in \S5.

\section{Observations}

NGC\,346 was observed once with the {\it Chandra} Observatory and
twice with the \x\ satellite. The {\it Chandra} data, taken on 
2001 May 15-16 for 100 ks, have been extensively analyzed in Papers 
I and II, and we will focus here only on the \x\ data. The first \x\ 
observation took place during revolution 0157, on Oct. 17 2000 for 
about 20 ks. It is a public dataset part of the \x\ Science Archive. 
The second observation, of length $\sim$30 ks, was obtained by our 
group during revolution 0357, on Nov. 20-21 2001. The EPIC MOS 
cameras \citep{tur01} were used in full frame mode during both 
revolutions, while the EPIC pn camera \citep{str01} was used in full frame 
mode during Rev. 0357 and in extended full frame mode during Rev. 0157. 
For both observations, a medium filter was added to reject optical light.

We used the Science Analysis Software (SAS) version 5.4.1. \citep{jan} 
to reduce the EPIC data. These data were first processed through the 
pipeline chains, and then filtered. For EPIC MOS, only events with a 
pattern between 0 and 12 and passing through the \#XMMEA\_EM filter 
(i.e. flag\&0x766b000=0) were considered. For EPIC pn, 
we kept events with flag$=$0 and a pattern between 0 and 4. 
The resulting exposure times were 19.6~ks (EPIC MOS) and 15.5~ks
(EPIC pn) for Rev. 0157, and 26.7~ks (EPIC MOS) and 21.7~ks
(EPIC pn) for Rev. 0357.

To search for contamination by low energy protons, we have further 
examined the light curve at high energies (PI$>$10000, 
approximately corresponding to energies $>$10~keV, and with 
pattern$=$0). Throughout both 
observations, a rather large high energy count rate (on average, 
0.25~cts~s$^{-1}$ for EPIC MOS and 1.25~cts~s$^{-1}$ for EPIC pn) 
was detected but since no clear flare is present, we decided to 
keep the whole exposure in both cases. Further analysis was performed 
using the SAS version 5.4.1. and the FTOOLS tasks \citep{bla}. The 
spectra were analyzed and fitted within XSPEC v11.0.1 \citep{ar96}.

\section{Source Properties}

The combined EPIC images of the NGC\,346 field of both \x\
datasets are shown in Fig. \ref{totalfield}. Several point sources 
can be spotted throughout the field and some of them are clearly 
seen to vary between the two observations. To create a catalog of 
these discrete X-ray sources, we applied for each observation the 
SAS detection meta-task {\it edetect\_chain} simultaneously to the
data from all three EPIC cameras. 
Due to the rather high background during these observations, we 
decided to restrict the energy range to 0.4$-$4.0~keV, and then created
images in the following three energy bands for input in {\it edetect\_chain}: 
S=0.4$-$1.0~keV, M=1.0$-$2.0~keV and H=2.0$-$4.0~keV. We eliminated false 
detections due to structures in the extended emission near \hd, gaps,
or remaining bad pixels and bad columns. We also rejected sources with 
a small detection likelihood (i.e.\ with a logarithmic likelihood in one
detector $\ln p_i \ll 5$ and/or a combined likelihood for the three detectors
$\ln p_i<30$). We have compared the lists found in each case
and finally constructed a definitive list of 51 sources\footnote{
The \x\ data from Rev. 0157 were used in the creation of the 1XMM 
catalog \citep{ssc}, which was released recently. However, this catalog
does not seem very accurate in this field, probably as a result of
the difficulties of the automated source detection algorithms in
complicated fields like this one \citep[see also ][]{gos}.
The 1XMM source list may have been created using a very low detection threshold
or maybe its algorithm was somewhat confused by  the presence of extended
emissions: most of the 108 1XMM sources in the field simply do not appear 
in the combined EPIC images. If we restrict ourselves to the sources
securely detected in the three cameras (i.e. with $Qflag=4$), only 20 sources 
remain, but some obvious, bright X-ray sources are not part of this 
restricted list and in addition, some of the ``secure detections'' are 
actually  the same source detected twice. A good source list for this
field thus still needed to be made.
} which are 
presented in Table \ref{crate} by increasing Right Ascension (RA). Note 
that 29 of these were already detected in our {\it Chandra} observation: 
for these sources, we will use the {\it Chandra} source number from 
Paper II, which will be quoted as [NHS03] \# throughout this paper, 
whereas for the sources specific to the \x\ observations, we will use 
a letter as identifier.
Most of the undetected {\it Chandra} sources have an ACIS-I count rate
smaller than $8\times10^{-4}$~cts~s$^{-1}$ (see also \S 3.3).
This is not surprising since the smallest \x\ count rates detected
in our observations are about $10^{-3}$~cts~s$^{-1}$ for EPIC MOS
and $2\times10^{-3}$~cts~s$^{-1}$ for EPIC pn. This detection threshold
is roughly an order of magnitude larger than for {\it Chandra} 
($10^{-4}$~cts~s$^{-1}$). The smaller exposure time, combined with the 
loss of resolution due to the larger Point Spread Function (PSF) of 
\x\ and a rather high background explain why we detect only one third 
of the {\it Chandra} sources. 

The count rates of the sources in each instrument are also presented in 
Table \ref{crate}. A missing count rate indicates a source (at least 
partly) in gaps or out of the field-of-view (FOV) for the instrument 
considered. If the task {\it edetect\_chain} provides vignetting- and 
background-corrected count rates, an additional correction is needed.
Following the User Support Group, it actually appears that the SAS task 
{\it eexpmap}, used by the meta-task {\it edetect\_chain},
 does not consider all the necessary exposure information. In 
particular, it neglects out-of-time (OOT) and deadtime corrections. If 
this is negligible for the EPIC MOS data, not correcting for OOT events 
in the EPIC pn data implies an overestimation of the exposure time (and 
thus an underestimation of the count rates) by 6.3\% for the full frame 
mode (Rev. 0357) and 2.3\% for the extended full frame mode (Rev. 0157). 
We thus have corrected the pn count-rates provided by {\it edetect\_chain} 
by this amount, and the result is shown in Table \ref{crate}. The 
presented hardness ratios HR1 and HR2 have been calculated for the pn 
data as (M$-$S)/(M+S) and (H$-$M)/(H+M), respectively. 
The last column of Table \ref{crate} yields information on the nature 
of the X-ray source \citep[e.g. ][]{ha04} or of its counterpart (Paper II
and Table \ref{wfi}). 
In addition to these point sources and the supernova remnant (SNR) 
near \hd, we note the possible presence of extended emission north of 
source E and near position 
0$^h$58$^m$25$^s$ $-$72$^{\circ}$00\arcmin40\arcsec, but their 
low surface brightness did not permit to derive any precise physical
information on these extended emissions.

\subsection{Source spectra and luminosities}

For the 12 brightest sources, we extracted the sources' spectra 
in circles of minimum radius 25\arcsec, and chose annuli surrounding 
the sources or nearby circles for background subtraction.
Using the SAS's {\it rmfgen} and {\it arfgen} tools we then generated 
the response matrix file and  ancillary reponse file, necessary to any 
spectral analysis. 
The spectra were finally binned to obtain at least 10 cts 
per bin, and were then analyzed within XSPEC. Bins with energies 
below 0.4~keV or above 10.0~keV were discarded. Since their signal-to-noise
was generally rather low, the spectra were fitted with simple models, i.e.
an absorbed power-law or an absorbed $mekal$ model. As fits to the MOS 
and pn spectra generally gave similar results, within the errors, we 
chose to fit simultaneously all EPIC data available for a particular 
observation to get the smallest errors. The results of these fits are 
presented in Table \ref{spec}, but $mekal$ models \citep{ka92} 
with unconstrained and unrealistically high temperatures (i.e. 
$kT>10.0$~keV) were not included in that table. We caution that the 
abundance specific to the SMC, $Z=0.1\,Z_{\odot}$, was used only 
for the $mekal$ models, not for the absorbing column 
\citep[like e.g. in ][]{ha04}. Absorbing columns larger than 
$\sim4\times10^{20}$~cm$^{-2}$ may indicate absorption 
in addition to the Galactic column. If this excess absorption is caused
by low metallicity SMC material, the SMC absorbing column can be 
derived by multiplying the excess column by a factor of 10.
The spectral properties of eight of our sources were also determined 
by two other teams \citep{sa03,maj}. The properties derived here are 
generally in good agreement with their values, since the confidence 
intervals overlap for all sources except \n\,75. In this last case, 
the spectral changes might actually be real, since the source is a 
varying X-ray binary (see \S\S 3.3.2 and 4.3).

\subsection{Counterparts}

For counterparts to the 29 \x\ sources which are in common with 
the {\it Chandra} observation, we refer the 
reader to the extensive discussion of Paper II, which we will not repeat 
here. To search for optical counterparts to the remaining X-ray sources, 
we have followed a procedure similar to that of Paper II. We have first
cross-correlated the source list with the public databases (USNO B1.0, 
GSC 2.2, 2MASS All Sky Survey) and then with our Wide Field Imager (WFI) 
data. Before correlating our source list with these optical databases, we have 
determined the mean shifts between {\it Chandra} and each \x\ observation. 
Since there exists a shift of +1\farcs6 in RA and $-$1\farcs2 in DEC between 
actual world coordinates and {\it Chandra}'s (see Paper II), we applied a 
shift of +0.26s in RA to coordinates of Rev. 0357 and +2\arcsec\ 
in DEC for Rev. 0157. All objects within 5\arcsec\ - the PSF width of 
\x\ - of the X-ray sources were considered as possible counterparts and
are listed in Table \ref{wfi}. Fig. \ref{colmag} presents a color-magnitude
diagram similar to Fig. 6 of Paper II. The majority of the sources are within 
the main sequence of NGC \,346 but the counterparts of sources I 
\& N appear as slightly evolved early-type stars and that a few counterparts 
(A,B,Q) apparently belong to a second population of the SMC, that is
not physically associated with the NGC\,346 cluster itself. 
We have also compared our source list to the {\it ROSAT} and {\it ASCA}
catalogs of SMC X-ray sources (Kahabka et al.\ 1999; Haberl et al.\ 2000,
hereafter [HFP2000]; Sasaki, Haberl, \& Pietsch 2000, hereafter 
[SHP2000]; Yokogawa et al.\ 2000, hereafter [YIT2000]). Thirteen of the 
22 \x\ sources were previously detected and we noted in Table \ref{wfi}
their \ro\ or {\it ASCA} identification. 

\subsection{Source Variability}

\subsubsection{Short-Term Variability}
We have also analyzed the lightcurves of these 12 brightest sources. 
Using the same regions as in \S3.1, we created lightcurves for each 
source and associated background region. The sources' lightcurves were 
then background-subtracted, corrected for good time intervals, and 
analyzed with our own software. Using Kolmogorov-Smirnov, $\chi^2$, 
and modified probability of variability (see Sana et 
al. 2004) tests, we found no source significantly 
variable in all three EPIC cameras during a single observation. 

\subsubsection{Long-Term Variability}
In Paper I and II, we have extensively analyzed
the sources detected by {\it Chandra}. Among other properties, we have 
determined the detailed spectral characteristics of 15 ACIS-I sources. 
Four out of these 15 sources were bright enough to provide a valuable EPIC 
spectra that can be compared with our previous results (see Paper II 
and Table \ref{spec}). All EPIC spectra present a lower absorption
column, compared to {\it Chandra}'s. This systematic difference
could most probably be attributed to calibration problems at low energies,
since the calibrations of both satellites are still ongoing
(in particular due to the recently discovered degradation of the ACIS 
low energy quantum efficiency).
In addition, these sources also present a clearly variable X-ray luminosity 
over the three observations:
\begin{itemize}
\item \n\ 4 has approximately the same flux in the {\it Chandra} data
as in the \x\ Rev. 0157 data, but it increased by 70\% in the
last \x\ observation.
\item \n\ 6 presents similar characteristics in the {\it Chandra} 
data and the first \x\ observation, but the power law steepens and the 
flux has dramatically decreased by a factor of $\sim$25 in the last \x\ observation.
\item \n\ 60 has a flux reduced by at least a factor 3 in the {\it 
Chandra} observation, compared to the \x\ ones. With its rather large
absorbing column and large power law energy slope $\Gamma$, it is 
considered in \citet{sa03} as a possible AGN candidate.
\item \n\ 69 presents only a small increase of flux ($\sim$20\%)
in the last \x\ observation, and marginal variations of the power law's
exponent.
\end{itemize} 

The spectral characteristics of ten additional {\it Chandra} sources 
are known, but they present too few counts in the  \x\ observations
to give any meaningful spectrum. To detect possible variations for these
sources, we used the spectral information from Paper II and PIMMS\footnote{
Available at http://heasarc.gsfc.nasa.gov/Tools/w3pimms.html. 
The closest $kT$ and $Z=0.2 Z_{\odot}$ (the lowest available) were used 
for the mekal models.} to predict the \x\ EPIC count rates, which we 
have then compared directly to the observed ones, since the output
of the task {\it edetect\_chain} is an equivalent on-axis count rate. By this 
procedure, we detected no flux variation for \n\ 1, 19, 20, 29, 
30, and 70, but four additional varying sources were also found:
\begin{itemize}
\item Using the \ch\ data, we expect \n\ 10 and 71 to be brighter by
40\% (EPIC MOS) and 200\% (EPIC pn) for the former, and 10\% (EPIC MOS)
and 50\% (EPIC pn) for the latter.
\item The expected EPIC pn count rate of \n\ 24 is a factor $\sim$4 
larger than the value observed at Rev. 0157. However, this cannot be 
confirmed by the EPIC MOS data of Rev. 0157, since the source falls 
in a detector gap at that time.
\item \n\ 46 possesses an \x\ count rate double of what is expected
on the basis of the {\it Chandra} data.
\end{itemize} 

We also note that the {\it Chandra} sources \n\ 23, 37, 47, and 61 
are detected with \x\, although they possessed an ACIS-I count rate 
smaller than $8\times10^{-4}$~cts~s$^{-1}$, i.e. so small that it 
should have prevented their detection by \x. 
On the contrary, \n\ 45 and 49 remain undetected (with \x\ count rates
$<1.5\times10^{-3}$~cts~s$^{-1}$ for EPIC MOS and 
$<3\times10^{-3}$~cts~s$^{-1}$ for EPIC pn) whereas they were 
expected to show up on the basis of their relatively large {\it Chandra} 
count rate (9.8 and 11.0 $\times10^{-4}$~cts~s$^{-1}$, respectively).
These six sources may be added to the list of variable X-ray sources.

In addition, for the sources out of {\it Chandra} ACIS-I FOV, we can
compare directly the count rates and/or spectral properties between both 
\x\ observations and we found 7 additional varying sources:
\begin{itemize}
\item \n\ 75 has undergone a dramatic increase of luminosity by a factor 
$\sim$23 in the last \x\ observation. As for \n\ 6, the power law 
steepened when the flux was lowest (see Fig. \ref{spec75}).
\item Source G has a slightly lower flux (a 15\% decrease) in the last 
\x\ observation and the flux of Source I is larger by 60\% in 
the first \x\ observation. Regarding the nature of these sources,
we may note that with its large absorbing column and $\Gamma$, 
Source G is a good AGN candidate, whereas Source I possesses 
a bright blue counterpart indicating a possible X-ray binary nature 
(see \S 4.3 for more details).
\item Source L exhibited a luminosity decrease by 75\% in the data from 
Rev. 0157.
\item The count rate of Source M has increased by 75\% in the last \x\ 
observation.
\item Source S has experienced an increase of 50\% in the power law exponent, 
associated with a decrease of the flux by a factor of 2 in the last
\x\ observation. This variation is similar to the behavior exhibited by 
\n\ 6 and 75.
\item The flux of Source V has doubled in the last \x\ observation. 
\end{itemize} 

There were also some 13 sources detected only in one of the two \x\ 
observations. However, most of these sources possess a very low count 
rate, close to the detection limit ($\sim10^{-3}$~cts~s$^{-1}$). Their 
absence could well be explained by a simple statistical fluctuation, and 
the upper limits derived in the observation where they are missing
are compatible with a constant count rate (see Table \ref{crate}). 
By comparing the observed count rates in one dataset to the
upper limits calculated for the other \x\ observation, we nevertheless
inferred count rate variations for Source N, of at least 20\% in EPIC MOS, 
and at least 170\% for EPIC pn. 
In addition, a large luminosity decrease is detected for source D. 
This source is at best marginally detected in the \x\ data of Rev. 
0357, while it is one of the brightest sources in the \x\ observation 
made at Rev. 0157: the count rate of this source changes by a factor 
of at least 25 between the two \x\ observations. 
In addition to finding pulses with a period of 280.4s, \citet{ts99} already 
discovered dramatic flux variations for this source by comparing \ro\ 
and {\it ASCA} observations, making Source D one of the most interesting 
X-ray binaries of this field. 

Comparing with older X-ray catalogs, we also note that the \ro\ sources 
[HFP2000] 103, 173, 185, 186, 207 and [KPF99] 157 are missing in the \x\ 
observation. Since these six \ro\ sources were not particularly faint 
\ro\ sources, their non-detection suggests a strongly variable nature. 
Using the SAS task $esensmap$ for a likelihood of 10, we derived upper 
limits on the \x\ count rate of these sources of 
$1.5-2.5\times10^{-3}$~cts~s$^{-1}$ for EPIC MOS and 
$3-4\times10^{-3}$~cts~s$^{-1}$ for EPIC pn. We note 
however that [HFP2000] 186 was already undetected in the {\it Chandra} 
observation (see Paper II), i.e. its ACIS-I count rate was lower than 
$10^{-4}$~cts~s$^{-1}$. In addition, the ACIS-I count rates of the \ro\ 
sources [HFP2000] 173 and 185 (=\n\ 11 and 16 in Paper II) were 
very low, about $2\times10^{-4}$~cts~s$^{-1}$, in our \ch\ 
data. If these three sources did not change since then, they are well 
below the detection threshold of our \x\ observations. 

\section{Comments on individual sources}

\subsection{\hd}

We reported the first detection of \hd\ at X-ray energies in Paper I. 
But this peculiar star was also observed twice by \x\ and in fact,
the \x\ data of Rev. 0157, which were taken before the \ch\ 
observation, constitute the first actual detection of \hd\ in X-rays. 
When {\it Chandra} caught the system at orbital phase $\phi=0.23-0.30$ 
(using the ephemeris of Sterken \& Breysacher 1997), 
the first \x\ observation sampled phase $\phi=0.36-0.38$ (eclipse 
of star A, the eruptor, by star B), while the data taken during Rev. 
0357 show \hd\ at $\phi=0.09-0.11$ (periastron). However, the poorer 
resolution of \x\ compared to {\it Chandra}'s renders difficult the 
disentangling of the source from the surrounding SNR (see Fig. \ref{zoom}), 
especially since the flux and spectral properties of this SNR change
spatially (see Paper I). 

To overcome these problems and provide hints of the intrinsic variations 
of the star, we chose three circular regions of 15\arcsec\ radius, which 
we will called `hd', `snr', and `bkgd'. Region `hd' is centered on \hd, 
but also sampled part of the SNR. Region `snr' is close-by area containing 
only contributions from the SNR. It is centered on {\it Chandra} coordinates 
0$^h$59$^m$32.5$^s$ $-$72$^{\circ}$10\arcmin26\arcsec. Region `bkgd'
is a nearby background region situated at {\it Chandra} coordinates 
0$^h$59$^m$45$^s$ $-$72$^{\circ}$11\arcmin00\arcsec. To eliminate the 
large background and sensitivity variations that may exist between our 
observations, we computed the ratio (hd$-$bkgd)/(snr$-$bkgd) in four energy 
ranges for each observation and for each instrument. 
We present in Fig. \ref{hd5980} the evolution with phase of that ratio
for the ranges 0.4$-$1.0~keV, 1.0$-$2.0~keV and 0.4$-$10.0~keV. Note 
that the results from the last chosen energy range, 2.0$-$4.0~keV, 
are too noisy and thus not very reliable: they are not displayed in 
Fig. \ref{hd5980}. 
There is no significant change between the \x\ data from Rev. 0357
and the $Chandra$ observation, but the X-ray luminosity of 
\hd\ has clearly increased during the first \x\ observation. 
This variation is particularly well seen in the harder range, which 
is not surprising since the emission from the SNR dominates at low 
energies.

The intrinsic X-ray emission of a colliding-wind system like \hd\ 
is expected to vary as 1/$D$ ($D$ being the separation between the components)
and it should thus be larger at periastron. We do not observe such a 
variation, but the effects of a varying absorption column might explain 
the observed changes. However, \hd\ is known to present
secular variations in the optical domain: with only three observations, it
is not yet possible to disentangle the possible long-term variations of
the X-ray luminosity (for example related to the 1994 eruption) from
systematic changes linked to the orbital motion. A more complete 
discussion about \hd\ variations will be presented by Flores et al. 
(in preparation), but we already note that a monitoring of \hd\ with the 
{\it Chandra} satellite (the only one easily capable of disentangling 
\hd\ from the SNR) would be of utmost interest to better constrain the 
system's physical characteristics.

\subsection{SNR\,0056$-$72.5}

Apart from the SNR surrounding \hd, a second supernova
remnant is present in the \x\ FOV: source E. As could be expected 
from such an object, this source did not vary between either of the \x\ 
observations. It is correlated with an extended non-thermal radio 
source discovered by \citet{ye91}. This radio source has a size of 
100\arcsec$\times$80\arcsec\ (or 160\arcsec$\times$160\arcsec\ 
if we use the lower contours of Ye et al.).
The X-ray source corresponds exactly with the peak of the radio 
source and, as far as the \x\ resolution can conclude\footnote{The 
PSF width of \x, $\sim$5\arcsec, corresponds to a rather large linear 
size in the SMC (1.4~pc).}, the \x\ source is point-like, i.e. not extended 
as in the radio range. However, its X-ray luminosity 
($\sim9\times10^{34}$ erg s$^{-1}$) is quite large for a putative 
isolated neutron star born in the supernova event \citep{hab}. 
The source also presents a rather hard spectra, that was well fitted 
by a power-law of energy slope $\Gamma\sim1.7$ but not by a thermal 
model ($kT$ of 4 to 50~keV). Finally, some faint extended emission 
may also be present to the north of the X-ray source, and it correlates
well with an extension of the source at radio wavelengths.

\subsection{X-ray Binaries}

Six X-ray binary candidates were proposed in our \x\ 
FOV by \citet{ha00}: \n\ 4, 6 \& 70; and Sources D, S \& V. Their status 
as XRBs is now confirmed by a large body of evidence. The X-ray 
sources \n\ 6 \& 70, and D \& S were found to pulsate 
\citep{ts99,sa03,mac}\footnote{Note that \citet{mac} failed to detect the 
variations of \n\ 6 over the {\it Chandra} exposure, which we reported 
in Paper II.}. All sources have also been found to vary on short- or 
long- timescales in \ro, {\it Chandra} and/or \x\ data 
\citep[Paper II; this work; ][]{ts99,sa03}. Moreover, of these 
six sources, only \n\ 70 appear to stay at a stable low luminosity 
level since 2000 Oct. (Rev. 0157).  In addition, all but \n\ 4 possess 
an emission-line, early-type counterpart (see e.g. Paper II). However, 
the candidate counterpart of Source V ([MA93] 1277, see Sasaki, Pietsch, 
\& Haberl 2003) lies rather far away from the X-ray source ($>$6\arcsec, 
a distance much larger than for the other XRB counterparts). Although the 
error on the position of this X-ray source should be larger since it is 
quite far off-axis ($\sim$11\arcmin), this large distance might eventuelly
cast doubt on the identification of Source V with [MA93] 1277. 

A few other XRB candidates have recently been proposed by \citet{sa03}: 
\n\ 75, and Sources I \& N. The varying character of these sources, 
confirmed by our new \x\ observation, and the presence of a bright 
blue counterpart left little doubt about their XRB nature. This is 
especially true for Source I, discovered recently to pulsate in the 
X-ray domain \citep{maj}.
In addition, we observe a clear variation of the flux of \n\ 69 
in the last \x\ observation. Since pulses were also detected 
for this source by \citet{lam}, its identification as an X-ray binary 
is extremely likely.

In Paper II, because of the estimated spectral types of their counterparts 
or due to their variable nature, we had also tentatively proposed a few 
new XRB candidates. Of these, \n\ 3, 9 and 34 were unfortunately too faint 
to be detected by \x, and \n\ 61 and 68 do not show any flux variation 
between the \x\ observations. However, \n\ 10 and 71 present
small variations: these sources should be considered as the best choices
for future studies of XRB candidates. 
Outside the {\it Chandra} FOV, another X-ray source may constitute an 
additional XRB candidate: Source U. We estimated a B spectral type 
for its counterpart (see Table \ref{wfi}) but no 
variations of the X-ray properties of the source were detected between 
our two \x\ observations.

Such a large number of XRBs in the SMC is not at all surprising 
\citep[e.g. ][]{ha00}, but a precise knowledge of their exact number and 
physical properties may enable us to better constrain their dependence 
on metallicity. The next step will be to monitor the counterparts of 
all these sources to find the physical parameters of these systems, 
and especially to determine the exact nature of the accretor.

Finally, the variability of the other sources described in \S3.3 is 
an interesting new discovery, that can give hints on the exact nature 
of these sources. They might be additional X-ray binaries, or varying 
extragalactic sources (AGNs). In this context, we may note that 
using the formalism of \citet{gia}, we expect about 20 extragalactic
sources in this field. Such sources should be well fitted by 
power-laws, but should also display a large absorbing column. 
The large slopes $\Gamma$ of \n\ 24, 46, \& 60 and of 
Sources G \& L may indicate that these sources are likely AGNs. 
Note however that amongst these, only Source G has a very large 
absorption column. Moreover, the known X-ray binary \n\ 4 was also 
fitted by a power-law of large energy slope $\Gamma$, and X-ray 
binaries containing a black hole can present large $\Gamma$ in 
their soft state (e.g. Cyg X-1). Long term monitoring of these 
sources and a precise study of their actual counterparts are needed, in 
order to better constrain their nature.

\section{Summary and conclusion}

In this third Paper on the X-ray emission from the NGC\,346 field, we 
have analyzed \x\ data taken six months before and after our {\it Chandra} 
observations. 51 sources were detected with \x, 29 of them being in
common with the {\it Chandra} data analyzed in Paper II. A comparison 
of the X-ray observations of the field has revealed the variations of 
21 of these 51 X-ray sources, 10 of them being known as X-ray 
binaries or XRB candidates. Another varying source is \hd, which 
appears brighter during the first \x\ observation. However, the exact
nature of these changes (secular or phase-locked variations ?) is not yet 
known, and requests additional X-ray data to be elucidated.

\acknowledgments

Y.N. acknowledges support from the PRODEX XMM-OM and Integral Projects, 
contract P5/36 `P\^ole d'Attraction Interuniversitaire' (Belgium).
This Paper utilizes public domain data from the 2MASS, USNO,
and the GSC; it has made use of the SIMBAD database, operated at CDS, 
Strasbourg (France) and NASA's ADS Abstract Service.
A.F. thanks the NASA/GSFC where part of this investigation was carried out.
The authors also acknowledge HEASARC for providing FTOOLS and XSPEC;
and the referee for useful comments.


\begin{figure}
\epsscale{0.45} 
\caption{The NGC\,346 field as seen by the EPIC instruments onboard 
\x. Left: data from Rev. 0157, Middle: data from Rev. 0357, 
Right: combined data from both revolutions with identification numbers 
(see Table~1 for a precise list of positions). Note that these figures 
were made by keeping the EPIC pn events with a pattern between 0 and 12 
and passing through the \#XMMEA\_EP filter.
\label{totalfield}}
\end{figure}

\begin{figure}
\epsscale{0.6} 
\caption{The Color-Magnitude Diagram of the counterparts listed in Table 
\ref{wfi}. The error bars correspond to the dispersion of the measured 
data. Counterparts with $V>$ 20~mag were not included since their photometry 
is very uncertain. The solid line shows an isochrone of 5~Myr for $Z$=0.004 
\citep{le01} transformed using a distance of 59~kpc and 
reddened with $R_V$=3.3 and $E(B-V)$ of 0.14~mag \citep{ma89}.
\label{colmag}}
\end{figure}

\begin{figure}
\epsscale{0.6} 
\caption{Unfolded spectra of \n\ 75 in the two \x\ observations, 
shown along with the best fit power law (see Table \ref{spec}).
The upper spectrum presents the EPIC MOS2 data of Rev. 0357, and
the lower one corresponds to the EPIC MOS1 data of Rev. 0157 (in the 
other cameras, the source was either in a gap, out of FOV or close to 
the edge of the FOV). In Rev. 0157, the luminosity in the 0.4$-$10~keV 
energy range has decreased by $>200$\%, 
while the power law slope has steepened from 1.1 to 3.9.
\label{spec75}}
\end{figure}

\begin{figure}
\epsscale{0.45} 
\caption{EPIC image of the SNR and \hd\ for both observations. The 
cluster NGC\,346 is just barely visible to the south-west of the SNR,
at position 0$^h$59$^m$04$^s$ $-$72$^{\circ}$10\arcmin35\arcsec (whereas
\hd\ is at 0$^h$59$^m$26$^s$ $-$72$^{\circ}$09\arcmin53\arcsec).
This figure uses the same event lists as Fig. 1.
\label{zoom}}
\end{figure}

\begin{figure}
\caption{Evolution of the diagnostic ratio for \hd\ (see text): open 
triangles show the {\it Chandra} data, filled squares the \x\ data 
from Rev. 0157 and filled circles the \x\ data from Rev. 
0357. The three points for each \x\ observation correspond to the 
ratio evaluated for the three EPIC cameras. The hard X-ray luminosity 
of \hd\ is clearly increasing in the \x\ observation from Rev. 0157. 
\label{hd5980}}
\end{figure}

\clearpage

\begin{table}
\setlength{\tabcolsep}{0.02in} 
\begin{center}
\tiny
\caption{\footnotesize Characteristics of the sources detected in the \x\ data. 
The first column gives the source number from Paper II, or a letter for 
the new sources. The second column gives the variability status of the 
source: `Y' for variable sources, `N' (resp. `n') if no variability was 
detected in the {\it Chandra} and \x\ observations (resp. only in the \x\ 
observations), and `?' for an uncertain status. The first column of each 
revolution yields the name of the source according to the conventions for 
serendipitous \x\ sources. The vignetting and background corrected EPIC-MOS 
and pn count rates in the 0.4$-$4.0~keV band are indicated for each instrument 
in 10$^{-3}$ cts s$^{-1}$, but the hardness ratios are given only for the pn 
data. The upper limits were determined using the task $esensmap$ (with 
$mlmin$=10). For sources affected by the gaps between the detectors or 
that are out of the FOV, we do not quote any count rate for the corresponding 
instrument. In the last column, we quote the information (nature, name,
pulse period) known on the X-ray source and/or its counterpart 
\citep[Paper II; this work; ][]{ha04,maj,lam}.
\label{crate}}\medskip 
\begin{tabular}{l c | l c c c r r } 
Src & V ?& \multicolumn{6}{c}{Rev. 0357} \\
& & & MOS1&  MOS2&  pn & HR1 & HR2 \\ 
\tableline
1 & N&XMMU J005652.2$-$721203 &  2.99$\pm$0.62  &  4.21$\pm$0.76  &  13.2$\pm$1.6  &$-$0.11$\pm$0.14 &$-$0.28$\pm$0.16\\
2 &n?&XMMU J005713.0$-$721041 &  1.42$\pm$0.49  &  1.08$\pm$0.47  &  3.89$\pm$0.99 &$-$0.26$\pm$0.27 &$-$0.11$\pm$0.40\\
4 & Y&XMMU J005732.5$-$721302 &  18.9$\pm$1.2   &  18.4$\pm$1.2   &  74.0$\pm$3.0  &$-$0.14$\pm$0.04 &$-$0.44$\pm$0.05\\
75& Y&XMMU J005735.6$-$721934 &                 &  68.5$\pm$2.6   &                &                 &                \\
5 &n?&XMMU J005741.8$-$720900 &  0.81$\pm$0.34  &  0.89$\pm$0.39  &  3.02$\pm$0.72 & 0.82$\pm$0.25   &$-$0.34$\pm$0.25\\
6 & Y&XMMU J005749.9$-$720756 &  2.86$\pm$0.52  &  4.40$\pm$0.63  &  9.11$\pm$1.08 & 0.66$\pm$0.14   &$-$0.20$\pm$0.12\\
10& Y&XMMU J005802.5$-$721205 &  0.83$\pm$0.32  &  1.54$\pm$0.40  &  1.46$\pm$0.63 &$-$0.32$\pm$0.50 &  0.06$\pm$0.60 \\
13& n&XMMU J005808.7$-$720826 &  1.27$\pm$0.37  &                 &  2.80$\pm$0.67 & 0.45$\pm$0.30   &$-$0.16$\pm$0.26\\
19& N&XMMU J005827.3$-$720500 &  1.39$\pm$0.38  &  1.82$\pm$0.47  &  2.97$\pm$0.62 & 1.00$\pm$0.46   &  0.22$\pm$0.18 \\
20& N&XMMU J005830.1$-$720842 &  1.66$\pm$0.34  &  1.53$\pm$0.39  &  5.39$\pm$0.73 & 0.09$\pm$0.15   &$-$0.38$\pm$0.16\\
23& Y&                        & $<$1.0          & $<$1.3          &                &                 &                \\ 
24& Y&XMMU J005836.8$-$720326 &  4.06$\pm$0.60  &  4.15$\pm$0.66  &                &                 &                \\
26&n?&XMMU J005839.7$-$720229 &  1.07$\pm$0.40  &  1.21$\pm$0.42  &  3.23$\pm$0.77 & 0.52$\pm$0.30   &$-$0.20$\pm$0.25\\
29& N&XMMU J005900.8$-$721329 &  3.03$\pm$0.42  &  3.36$\pm$0.47  &                &                 &                \\
30& N&XMMU J005903.1$-$721224 &  2.55$\pm$0.37  &  4.03$\pm$0.49  &                &                 &                \\
36& n&XMMU J005912.9$-$721620 &  1.13$\pm$0.70  &  0.25$\pm$0.37  &  4.62$\pm$0.81 &$-$0.11$\pm$0.17 &$-$0.64$\pm$0.27\\
37& Y&XMMU J005918.0$-$721112 &  0.52$\pm$0.24  &  0.55$\pm$0.27  &  3.15$\pm$0.68 &$-$0.17$\pm$0.25 &$-$0.13$\pm$0.26\\
40& n&XMMU J005925.3$-$721432 &  0.66$\pm$0.25  &  0.64$\pm$0.26  &  2.83$\pm$0.60 &$-$0.01$\pm$0.21 &$-$0.81$\pm$0.25\\
43& n&XMMU J005931.6$-$721417 &  0.72$\pm$0.25  &  1.08$\pm$0.32  &  1.76$\pm$0.51 &   1.00$\pm$1.37 & 0.54$\pm$0.26  \\
46& Y&XMMU J005935.0$-$720211 &  4.68$\pm$0.67  &  4.48$\pm$0.61  &                &                 &                \\
47& Y&XMMU J005935.4$-$720836 &  0.52$\pm$0.25  &  0.20$\pm$0.23  &  2.25$\pm$0.50 & 0.83$\pm$0.35   & 0.07$\pm$0.21  \\
55&n?&                        & $<$1.1          & $<$1.4          & $<$2.0         &                 &                \\ 
60& Y&XMMU J010015.5$-$720442 &  9.36$\pm$0.85  & 12.3$\pm$1.0    &                &                 &                \\
61& Y&XMMU J010017.2$-$721054 &  0.49$\pm$0.26  &  1.47$\pm$0.39  &                &                 &                \\
62& n&XMMU J010022.9$-$721130 &  1.01$\pm$0.34  &  1.62$\pm$0.39  &  4.02$\pm$0.70 & $-$0.20$\pm$0.17&$-$0.71$\pm$0.29\\
68& n&XMMU J010037.8$-$721314 &  0.77$\pm$0.30  &  1.66$\pm$0.44  &  1.30$\pm$0.53 & 1.00$\pm$0.86   & 0.22$\pm$0.37  \\
69& Y&XMMU J010042.9$-$721133 &                 &                 &  227.$\pm$4.   & 0.05$\pm$0.02   &$-$0.61$\pm$0.02\\
70& N\tablenotemark{a}&XMMU J010102.5$-$720659 &  2.29$\pm$0.51  &  4.58$\pm$0.68  &  6.19$\pm$0.88 &$-$0.05$\pm$0.16 &$-$0.12$\pm$0.18\\
71& Y&XMMU J010103.8$-$721007 &  1.33$\pm$0.43  &                 &  3.51$\pm$0.75 &$-$0.38$\pm$0.22 &$-$0.10$\pm$0.34\\
\tableline
A &n?&XMMU J005651.3$-$720804 &  2.15$\pm$0.63  &  1.64$\pm$0.62  &  3.66$\pm$1.11 & 1.00$\pm$0.37   &$-$0.04$\pm$0.29\\ 
B &n?&                        & $<$1.7          & $<$2.1          &                &                 &                \\ 
C & n&XMMU J005725.6$-$721646 &  1.38$\pm$0.49  &  1.78$\pm$0.54  &  3.72$\pm$1.04 & 0.50$\pm$0.28   &$-$0.81$\pm$0.26\\
D & Y&                        & $<$1.8          & $<$2.0          & $<$2.9         &                 &                \\ 
E & n&XMMU J005816.7$-$721806 &  15.8$\pm$1.1   &  17.9$\pm$1.2   &  50.2$\pm$2.5  & 0.23$\pm$0.05   &$-$0.45$\pm$0.05\\
F &n?&XMMU J005846.1$-$715809 &  1.60$\pm$0.63  &  0.94$\pm$0.51  &  4.60$\pm$1.24 &$-$0.31$\pm$0.30 &$-$0.02$\pm$0.43\\
G & Y&XMMU J005914.4$-$722231 &  5.58$\pm$0.84  &  6.55$\pm$0.92  &  11.2$\pm$1.6  & 0.59$\pm$0.18   &$-$0.03$\pm$0.15\\
H &n?&XMMU J005916.5$-$715630 &  2.84$\pm$0.84  &  2.63$\pm$0.79  &  4.79$\pm$1.32 & 0.00$\pm$0.40   & 0.28$\pm$0.31  \\
I & Y&XMMU J005921.0$-$722317 &  29.5$\pm$1.8   &  40.3$\pm$2.1   &                &                 &                \\
J & n&XMMU J005947.1$-$720059 &  1.23$\pm$0.46  &  2.55$\pm$0.63  &  6.61$\pm$0.99 & 0.37$\pm$0.17   &$-$0.39$\pm$0.16\\
K & n&XMMU J010004.6$-$715921 &  1.29$\pm$0.58  &  0.25$\pm$0.39  &  4.70$\pm$1.03 &$-$0.19$\pm$0.22 &$-$0.26$\pm$0.35\\
L & Y&XMMU J010005.9$-$715724 &  13.7$\pm$1.6   &                 &  30.2$\pm$2.5  &$-$0.10$\pm$0.09 &$-$0.55$\pm$0.11\\
M & Y&XMMU J010011.8$-$722013 &                 &  6.48$\pm$0.90  &  19.8$\pm$1.8  & 0.58$\pm$0.09   &$-$0.45$\pm$0.09\\
N & Y&                        & $<$1.8          & $<$2.1          & $<$3.2         &                 &                \\ 
O & n&XMMU J010049.2$-$720347 &  1.91$\pm$0.54  &  3.51$\pm$0.65  &  5.59$\pm$0.97 & 0.68$\pm$0.27   & 0.01$\pm$0.17  \\
P & n&XMMU J010049.6$-$721408 &  2.75$\pm$0.51  &  2.14$\pm$0.52  &  4.15$\pm$0.79 &$-$0.05$\pm$0.20 &$-$0.39$\pm$0.25\\
Q & ?&                        &                 &                 &                &                 &                \\
R & n&XMMU J010103.2$-$721534 &  0.92$\pm$0.37  &  1.53$\pm$0.51  &  2.27$\pm$0.72 & 0.77$\pm$0.46   & 0.05$\pm$0.31  \\
S & Y&XMMU J010120.4$-$721119 &  17.5$\pm$1.2   &  26.8$\pm$1.6   &  50.0$\pm$2.3  & 0.39$\pm$0.06   &$-$0.07$\pm$0.05\\
T & n&XMMU J010127.4$-$721305 &  1.78$\pm$0.48  &  1.60$\pm$0.57  &  2.04$\pm$0.85 &$-$0.04$\pm$0.48 &$-$0.00$\pm$0.53\\
U & n&XMMU J010133.1$-$721320 &  1.34$\pm$0.48  &  2.24$\pm$0.59  &  4.42$\pm$0.91 & 0.03$\pm$0.27   & 0.04$\pm$0.24  \\
V & Y&XMMU J010137.4$-$720420 &  14.3$\pm$1.3   &  19.3$\pm$1.6   &                &                 &                \\
\end{tabular}
\end{center}
\end{table}
\clearpage

\setcounter{table}{0}
\begin{table}
\setlength{\tabcolsep}{0.02in} 
\begin{center}
\tiny
\caption{\footnotesize Continued}\medskip 
\begin{tabular}{l c | l c c c r r| l} 
Src & V ?& \multicolumn{6}{c|}{Rev. 0157} & Remarks\\
& & & MOS1& MOS2&  pn & HR1 & HR2&\\ 
\tableline
1 & N& XMMU J005652.3$-$721206 &  3.38$\pm$0.74  &  3.04$\pm$0.68  &  7.95$\pm$1.50 &$-$0.01$\pm$0.20&$-$0.25$\pm$0.26&\\
2 &n?&                         & $<$2.0          & $<$2.0          & $<$3.5         &                &                &\\
4 & Y& XMMU J005732.6$-$721304 &  12.9$\pm$1.2   &  9.95$\pm$1.05  &  41.8$\pm$2.8  &$-$0.09$\pm$0.07&$-$0.47$\pm$0.08& XRB cand.\\
75& Y& XMMU J005736.5$-$721936 &  4.30$\pm$0.75  &                 &  15.0$\pm$2.1  & 0.66$\pm$0.16  &$-$0.24$\pm$0.14&XRB, $P$=565~s, [MA93]1020\\
5 &n?&                         & $<$1.7          & $<$1.7          & $<$3.1         &                &                &\\
6 & Y& XMMU J005750.3$-$720758 &  51.4$\pm$2.3   &                 &  120.$\pm$4.   & 0.51$\pm$0.04  & 0.09$\pm$0.04  &XRB, $P$=152.3~s, [MA93]1038\\
10& Y& XMMU J005802.7$-$721206 &  1.35$\pm$0.40  &  0.94$\pm$0.38  &  3.02$\pm$0.79 & 0.83$\pm$0.33  &$-$0.00$\pm$0.26& B star\\
13& n& XMMU J005809.2$-$720826 &  0.90$\pm$0.42  &  0.44$\pm$0.36  &  3.07$\pm$0.89 &$-$0.12$\pm$0.31&$-$0.22$\pm$0.43&\\
19& N& XMMU J005828.0$-$720501 &  1.64$\pm$0.48  &                 &  4.10$\pm$0.77 & 0.96$\pm$0.14  &$-$0.22$\pm$0.19&\\
20& N& XMMU J005830.1$-$720844 &  2.17$\pm$0.42  &  1.78$\pm$0.42  &  5.78$\pm$0.88 & 0.45$\pm$0.19  &$-$0.19$\pm$0.16&\\
23& Y& XMMU J005831.7$-$720953 &  0.85$\pm$0.33  &  0.73$\pm$0.27  &  1.81$\pm$0.61 & 0.73$\pm$0.52  & 0.19$\pm$0.33  &\\ 
24& Y& XMMU J005838.1$-$720324 &                 &                 &  3.35$\pm$0.87 & 0.61$\pm$0.22  &$-$1.00$\pm$0.36&\\
26&n?&                         & $<$2.1          &                 &                &                &                &\\
29& N& XMMU J005901.1$-$721331 &  2.95$\pm$0.45  &  4.17$\pm$0.53  &  6.36$\pm$0.94 & 1.00$\pm$0.18  &$-$0.11$\pm$0.14&A star\\
30& N& XMMU J005903.3$-$721226 &  2.72$\pm$0.40  &  3.15$\pm$0.49  &                &                &                &\\
36& n& XMMU J005913.4$-$721620 &  1.55$\pm$0.39  &  1.04$\pm$0.34  &                &                &                &\\
37& Y&                         & $<$1.3          & $<$1.2          & $<$2.6         &                &                &\\
40& n& XMMU J005925.8$-$721433 &  0.61$\pm$0.26  &  0.23$\pm$0.23  &  2.07$\pm$0.57 & 1.00$\pm$0.22  &$-$0.59$\pm$0.30&\\
43& n& XMMU J005931.5$-$721419 &  0.73$\pm$0.27  &  1.16$\pm$0.30  &                &                &                &\\
46& Y& XMMU J005934.9$-$720213 &  5.03$\pm$0.81  &  3.13$\pm$0.64  &  8.84$\pm$1.26 &$-$0.05$\pm$0.15&$-$0.46$\pm$0.18&\\
47& Y& XMMU J005935.2$-$720837 &  0.38$\pm$0.27  &  0.98$\pm$0.28  &  0.86$\pm$0.44 & 1.00$\pm$0.31  &$-$0.53$\pm$0.48&\\
55&n?& XMMU J005952.8$-$721533 &  0.61$\pm$0.31  &  0.84$\pm$0.33  &  1.88$\pm$0.68 & 0.34$\pm$0.58  & 0.19$\pm$0.36  &\\
60& Y& XMMU J010015.7$-$720444 &  8.17$\pm$0.93  &  7.78$\pm$0.90  &  20.7$\pm$1.6  & 0.32$\pm$0.08  &$-$0.46$\pm$0.08&XRB cand. or AGN ?\\
61& Y& XMMU J010017.3$-$721051 &  0.86$\pm$0.33  &  0.71$\pm$0.34  &  1.27$\pm$0.60 & 1.00$\pm$1.05  & 0.27$\pm$0.44  &B star\\
62& n& XMMU J010023.3$-$721131 &  1.65$\pm$0.45  &  1.44$\pm$0.39  &  6.08$\pm$0.96 &$-$0.41$\pm$0.15&$-$0.52$\pm$0.31&\\
68& n& XMMU J010036.7$-$721321 &  1.11$\pm$0.38  &  0.85$\pm$0.37  &  1.81$\pm$0.72 &$-$1.00$\pm$15.7& 1.00$\pm$0.43  &B star\\
69& Y& XMMU J010043.0$-$721135 &  68.0$\pm$2.4   &                 &  200.$\pm$5.   & 0.05$\pm$0.03  &$-$0.63$\pm$0.03&XRB, $P$=5.44~s, B star\\
70& N\tablenotemark{a}& XMMU J010103.1$-$720702 &  3.42$\pm$0.60  &  4.34$\pm$0.64  &  8.90$\pm$1.18 &$-$0.05$\pm$0.15&$-$0.16$\pm$0.17&XRB, $P$=304.5~s, [MA93]1240\\
71& Y& XMMU J010103.8$-$721007 &  1.32$\pm$0.47  &  1.94$\pm$0.55  &                &                &                &B star\\
\tableline
A &n?&                         & $<$2.6          & $<$2.5          & $<$4.3         &                &                &\\ 
B &n?& XMMU J005722.8$-$721759 &  1.36$\pm$0.60  &  1.53$\pm$0.59  &  5.45$\pm$1.26 & 0.91$\pm$0.25  &$-$0.22$\pm$0.23&\\ 
C &n & XMMU J005726.5$-$721649 &  1.53$\pm$0.52  &  1.64$\pm$0.60  &  3.85$\pm$1.31 &$-$0.14$\pm$0.37&$-$0.17$\pm$0.45&\\ 
D &Y & XMMU J005749.7$-$720238 &  49.6$\pm$2.9   &  39.4$\pm$2.2   &  86.2$\pm$3.8  & 0.19$\pm$0.05  &$-$0.08$\pm$0.05&XRB, $P$=280.4~s, [MA93]1036\\
E &n & XMMU J005816.8$-$721806 &  14.8$\pm$1.3   &  12.7$\pm$1.2   &  48.4$\pm$3.0  & 0.19$\pm$0.07  &$-$0.17$\pm$0.07& SNR 0056$-$72.5 \\
F &n?&                         & $<$2.8          &                 &                &                &                &\\
G &Y & XMMU J005914.2$-$722232 &  5.99$\pm$0.96  &  5.48$\pm$0.93  &                &                &                &\\
H &n?&                         &                 &                 & $<$4.6         &                &                &\\
I &Y & XMMU J005921.0$-$722318 &  44.1$\pm$2.4   &  48.0$\pm$2.6   &                &                &                &XRB, $P$=202~s\\
J &n & XMMU J005947.0$-$720057 &  2.58$\pm$0.71  &  2.53$\pm$0.69  &                &                &                &\\
K &n & XMMU J010005.1$-$715925 &  0.96$\pm$0.64  &  1.18$\pm$0.65  &                &                &                &\\
L &Y & XMMU J010005.8$-$715725 &                 &  6.50$\pm$1.34  &  19.5$\pm$2.4  & 0.02$\pm$0.13  &$-$0.48$\pm$0.15&\\
M &Y & XMMU J010012.0$-$722014 &  3.79$\pm$0.75  &                 &  10.5$\pm$1.5  & 0.27$\pm$0.18  &$-$0.07$\pm$0.16&\\
N &Y & XMMU J010030.2$-$722033 &  2.52$\pm$0.62  &  2.18$\pm$0.67  &  8.84$\pm$1.58 & 0.82$\pm$0.19  &$-$0.25$\pm$0.17&XRB cand., [MA93]1208\\ 
O &n & XMMU J010049.9$-$720351 &  2.55$\pm$0.58  &  4.04$\pm$0.70  &  5.02$\pm$1.06 & 0.62$\pm$0.23  &$-$0.41$\pm$0.22&\\
P &n & XMMU J010049.8$-$721410 &  1.56$\pm$0.45  &  3.38$\pm$0.59  &                &                &                &\\
Q &? & XMMU J010057.4$-$722231 &  4.19$\pm$0.89  &  6.15$\pm$1.06  &                &                &                &\\
R &n & XMMU J010103.2$-$721538 &  0.75$\pm$0.38  &  1.00$\pm$0.46  &  2.78$\pm$0.88 & 0.67$\pm$0.35  &$-$0.72$\pm$0.30&\\
S &Y & XMMU J010120.7$-$721120 &  32.5$\pm$1.9   &  32.5$\pm$2.0   &  83.3$\pm$3.6  & 0.24$\pm$0.05  & 0.06$\pm$0.05  &XRB, $P$=455~s, [MA93]1257\\
T &n & XMMU J010127.9$-$721305 &  0.53$\pm$0.41  &  1.26$\pm$0.51  &  5.94$\pm$1.24 &$-$0.13$\pm$0.22&$-$0.37$\pm$0.32&\\
U &n & XMMU J010133.2$-$721322 &  2.06$\pm$0.60  &  1.53$\pm$0.56  &  3.91$\pm$1.14 & 0.46$\pm$0.37  &$-$0.22$\pm$0.30& B star\\
V &Y & XMMU J010137.6$-$720421 &  10.1$\pm$1.2   &  10.7$\pm$1.2   &  34.2$\pm$2.5  & 0.10$\pm$0.08  &$-$0.26$\pm$0.09&XRB cand., [MA93]1277\\
\end{tabular}
\tablenotetext{a}{This source did not vary between our three datasets, but was observed by \ro\ to be much brighter (see Paper II). }
\end{center}
\end{table}
\clearpage
\normalsize 

\begin{sidewaystable}
\begin{center}
\tiny
\caption{Parameters of the spectral fits. $\Gamma$ is the photon index 
of the power law, and the quoted limits correspond to the 90\% confidence 
interval. Abundances are fixed at $0.1 Z_{\odot}$, except for \n\ 69
for which it was fitted ($Z_{0157}=0.0015_{0.0001}^{0.0039}Z_{\odot}$ and 
$Z_{0357}=0.0003_{0.0001}^{0.0014}Z_{\odot}$, to be compared with 
$Z_{Ch}=0.0022_{0.0001}^{0.0094}Z_{\odot}$, cfr. Paper II). Absorbed 
luminosities in the spectral range 0.4$-$10~keV 
are given for a distance of 59~kpc, and clearly unphysical models have 
been discarded. All available EPIC spectra were fitted 
simultaneously. For comparison, the 90\% confidence intervals and the 
luminosity of the {\it Chandra} data are presented in the last columns.
\label{spec}}\medskip 
\begin{tabular}{l c c c c | c c c c| c c c} 
\tableline
Src & \multicolumn{4}{c|}{\x\ Rev. 0157} & \multicolumn{4}{c|}{\x\ Rev. 0357}& \multicolumn{3}{c}{Chandra}\\
 & $N({\rm H})$ &  & $\chi^2_{\nu}$(dof) & $L_X^{abs}$  & $N({\rm H})$ & & $\chi^2_{\nu}$(dof) & $L_X^{abs}$  & $N({\rm H})$ & & $L_X^{abs}$ \\
 & (10$^{22}$ cm$^{-2}$) & ($kT$ in keV) &  & (10$^{34}$ & (10$^{22}$ cm$^{-2}$)&   ($kT$ in keV) & & (10$^{34}$ & (10$^{22}$ cm$^{-2}$)& ($kT$ in keV) & (10$^{34}$\\
 &  & &  & ergs s$^{-1}$) & & & &  ergs s$^{-1}$)&&&ergs s$^{-1}$)\\
\tableline
\vspace*{-0.cm}&&&&&&&&&&&\\
\n\ 4 mek. & 0.$_{0.}^{0.04}$ & $kT$=3.01$_{2.13}^{4.48}$ & 1.12 (73) & 3.68& 0.05$_{0.02}^{0.07}$ & $kT$=2.27$_{1.96}^{2.83}$ & 1.04 (161) & 6.50& 0.13$-$0.19 &$kT$=2.5$-$3.6& 4.21\\
\vspace*{-0.cm}&&&&&&&&&&&\\
\n\ 4  pow. & 0.08$_{0.01}^{0.17}$ & $\Gamma=2.20_{1.97}^{2.62}$ & 1.11 (73) & 3.97& 0.21$_{0.15}^{0.26}$ & $\Gamma=2.66_{2.37}^{2.89}$ & 0.99 (161) & 6.71& 0.24$-$0.32 & $\Gamma$=2.2$-$2.5 & 4.54\\
\vspace*{-0.cm}&&&&&&&&&&&\\
\n\ 6  pow. & 0.31$_{0.25}^{0.39}$ & $\Gamma=0.81_{0.76}^{0.89}$ & 0.94 (309) & 60.0& 0.49$_{0.25}^{1.01}$ & $\Gamma=1.53_{1.05}^{1.90}$ & 1.00 (67) & 2.54& 0.45$-$0.56 & $\Gamma$=0.80$-$0.93 & 55.4\\
\vspace*{-0.cm}&&&&&&&&&&&\\
\n\ 60  mek. & 0.04$_{0.}^{0.21}$ & $kT$=5.66$_{2.37}^{43.}$ & 1.15 (26) & 4.00& 0.06$_{0.}^{0.16}$ & $kT$=4.08$_{2.21}^{13.}$ & 1.42 (46) & 4.00& 0.33$-$0.58 &$kT$=2.46$-$8.63 & 1.15\\
\vspace*{-0.cm}&&&&&&&&&&&\\
\n\ 60  pow. & 0.12$_{0.}^{0.34}$ & $\Gamma=1.82_{1.29}^{2.57}$ & 1.15 (26) & 4.29& 0.18$_{0.05}^{0.35}$ & $\Gamma=2.17_{1.76}^{2.83}$ & 1.35 (46) & 4.05& 0.35$-$0.77 & $\Gamma$=1.71$-$2.46 & 1.27\\
\vspace*{-0.cm}&&&&&&&&&&&\\
\n\ 69  mek. & 0.23$_{0.20}^{0.26}$ & $kT$=0.98$_{0.90}^{1.10}$ & 0.95 (241) & 15.1& 0.20$_{0.17}^{0.23}$ & $kT$=1.17$_{1.04}^{1.30}$ & 0.74 (253) & 18.0& 0.38$-$0.44 &$kT$=0.99$-$1.11 & 14.8\\
\vspace*{-0.cm}&&&&&&&&&&&\\
\n\ 69  pow. & 0.44$_{0.40}^{0.48}$ & $\Gamma=3.62_{3.45}^{3.81}$ & 1.06 (242) & 15.7& 0.39$_{0.36}^{0.43}$ & $\Gamma=3.34_{3.17}^{3.52}$ & 0.76 (254) & 19.1& 0.67$-$0.75 & $\Gamma$=3.71$-$3.93 & 13.5\\
\vspace*{-0.cm}&&&&&&&&&&&\\
\n\ 75  pow. & 2.61$_{0.84}^{5.23}$ & $\Gamma=3.90_{1.85}^{6.39}$ & 0.80 (10) & 2.95& 1.73$_{1.21}^{2.44}$ & $\Gamma=1.14_{0.83}^{1.51}$ & 0.86 (75) & 69.3& & & \\
\vspace*{-0.cm}&&&&&&&&&&&\\
\tableline
\vspace*{-0.cm}&&&&&&&&&&&\\
D pow. & 0.05$_{0.}^{0.21}$ & $\Gamma=0.79_{0.57}^{1.03}$ & 0.78 (46) & 44.1& & & & & & & \\
\vspace*{-0.cm}&&&&&&&&&&&\\
E pow. & 0.17$_{0.10}^{0.26}$ & $\Gamma=1.53_{1.35}^{1.75}$ & 0.95 (105) & 9.26& 0.19$_{0.09}^{0.26}$ & $\Gamma=1.81_{1.55}^{2.12}$ & 0.86 (70) & 9.07& & & \\
\vspace*{-0.cm}&&&&&&&&&&&\\
G pow. & 1.54$_{0.60}^{3.11}$ & $\Gamma=2.70_{1.55}^{4.59}$ & 1.07 (13) & 3.41& 0.95$_{0.31}^{1.99}$ & $\Gamma=2.32_{1.28}^{3.88}$ & 1.42 (20) & 2.94& & & \\
\vspace*{-0.cm}&&&&&&&&&&&\\
I pow. & 0.18$_{0.09}^{0.30}$ & $\Gamma=1.28_{1.09}^{1.48}$ & 0.71 (75) & 34.4& 0.37$_{0.25}^{0.52}$ & $\Gamma=1.49_{1.26}^{1.75}$ & 0.99 (68) & 21.0& & & \\
\vspace*{-0.cm}&&&&&&&&&&&\\
L mek. & 0.06$_{0.}^{0.16}$ & $kT$=1.96$_{1.31}^{5.22}$ & 1.06 (36) & 1.87& 0.07$_{0.}^{0.17}$ & $kT$=3.09$_{1.63}^{13.}$ & 1.11 (51) & 3.22& & & \\
\vspace*{-0.cm}&&&&&&&&&&&\\
L pow. & 0.23$_{0.09}^{0.43}$ & $\Gamma=2.84_{1.94}^{4.53}$ & 1.05 (36) & 1.91& 0.18$_{0.05}^{0.27}$ & $\Gamma$=2.24$_{1.88}^{3.14}$ & 1.09 (51) & 3.47& & & \\
\vspace*{-0.cm}&&&&&&&&&&&\\
S pow. & 0.11$_{0.04}^{0.20}$ & $\Gamma=0.60_{0.50}^{0.71}$ & 1.23 (194) & 38.3& 0.18$_{0.10}^{0.29}$ & $\Gamma=0.90_{0.79}^{1.05}$ & 0.99 (172) & 18.8& & & \\
\vspace*{-0.cm}&&&&&&&&&&&\\
V pow. & 0.20$_{0.05}^{0.38}$ & $\Gamma=2.10_{1.72}^{2.77}$ & 0.93 (58) & 4.02& 0.09$_{0.}^{0.25}$ & $\Gamma=1.67_{1.25}^{2.07}$ & 1.19 (40) & 7.81& & & \\
\vspace*{-0.cm}&&&&&&&&&&&\\
\tableline
\end{tabular}
\end{center}
\end{sidewaystable}
\clearpage

\begin{table}
\begin{center}
\scriptsize
\caption{Counterparts of the X-ray sources. The second column gives the 
\ro\ or {\it ASCA} name of the source (y=[YIT2000], h=[HFP2000], s=[SHP2000]), 
while the following columns present the WFI photometry of the optical 
counterparts, with the separation between the X-ray sources and their 
counterparts. The error quoted in the $\sigma_{V}$ column represents 
the dispersion of the measured data. If the counterpart is cataloged, the 
identifier, starting by S for GSC 2.2, U for USNO-B1.0, or 2M for 2MASS 
all Sky Survey, is given in the last column of the Table. When it 
was possible, an estimation of the spectral type of the counterpart 
was made (in italics, see last column), assuming that it belongs to 
NGC\,346 ($E(B-V)$=0.14~mag, d=59~kpc).
\label{wfi}}\medskip
\begin{tabular}{l r r r r r r c c c l} 
\tableline\tableline
Src & \ro & $V$ & $B-V$ & $U-B$ & $V-R$ & $R-I$ & $\sigma_{V}$ & d157& d357& Remarks\\
& or {\it ASCA}&(mag)&(mag)&(mag)&(mag)&(mag)& (mag)& (\arcsec)& (\arcsec)&\\ 
\tableline
A &        & 17.86 &  0.86 &  1.06 &  0.50 &  0.48 &0.03 &      &  2.3& U0178-0040763\\
  &        &       &       &       &       &       &     &      &     & =2M00565129$-$7208034\\
B &        & 18.46 &  0.79 &  0.15 &  0.45 &  0.34 &0.03 &  3.8 &     & S01020202023\\
C &        & 20.30 &$-$0.30&       &  0.45 &       &0.20 &  1.3 & 3.5 &\\
  &        & 19.61 &  0.03 &       &$-$0.01&       &0.10 &  3.2 & 3.9 &$B9V?$\\
D &y20,h114& 15.62 &$-$0.07&$-$0.96&  0.09 &$-$0.02&0.02 &  1.2 &     & $lateO,earlyBV$,[MA93]1036\tablenotemark{a}\\
E &y21,h194& 20.00 &  0.10 &       &$-$0.08&       &0.20 &  3.3 & 4.7 &SNR\,0056$-$72.5\\
F &        & 18.06 &$-$0.23&$-$0.76&$-$0.10&$-$0.05&0.04 &      &  4.0&\\
G &s80     & 19.37 &  0.92 &$-$0.77&  0.59 &  0.79 &0.10 &  0.5 & 2.4 &\\
H &        & 19.78 &$-$0.13&$-$0.31&  0.05 &  0.55 & 0.23&      &  3.0&\\
I &h218    & 14.79 &  0.11 &$-$1.09&$-$0.02&$-$0.05& 0.07&  0.6 & 2.2 & S010231088040=U0176-0044581\\
  &        &       &       &       &       &       &     &      &     & =2M00592103$-$7223173\\
J &y23,h102& 19.45 &  0.78 &       &  0.48 &  0.66 & 0.08&  2.0 & 2.3 &\\
K &        & 19.87 &  0.31 &       &  0.34 &  0.28 &0.05 &  2.0 & 1.7 &\\
  &        & 19.46 &  0.50 &       &  0.39 &  0.25 &0.08 &  2.4 & 4.0 &\\
L &h91     & 19.18 &  0.67 &$-$0.47&  0.19 &  0.63 & 0.06&  0.6 & 2.5 &\\
M &h204    & \\
N &        & 14.71 &  0.13 &$-$1.06&  0.10 &  0.01 & 0.02&  2.0 &     & [MA93]1208\tablenotemark{b}\\
O &s91     & 19.12 &$-$0.14&$-$0.54&$-$0.18&       &0.05 &  5.0 & 3.0 &\\
P &h177    & 19.95 &  0.07 &$-$0.36&$-$0.39&       &0.19 &  3.2 & 3.5 &\\
Q &h213    & 18.06 &  1.13 &  0.26 &  0.53 &  0.54 & 0.06&  3.5 &     & U0176-0047908\\
S &h159    & 15.63 &$-$0.11&$-$1.06&  0.02 &$-$0.03& 0.02&  0.7 & 1.2 &$lateO,earlyBV$,[MA93]1257\tablenotemark{c}\\
T &y27     & 21.26 &$-$0.17&       &$-$0.12&  2.84 &0.27 &  3.0 & 3.0 &\\
U &        & 18.66 &$-$0.02&$-$0.48&$-$0.03&  0.09 & 0.06&  3.9 & 3.5 &$B5-7V$\\
V &h121    & 16.22 &$-$0.07&$-$0.90&  0.05 &$-$0.02& 0.04&  5.7 & 7.7 &$B1V$,[MA93]1277\tablenotemark{d}\\
\tableline
\end{tabular}
\tablenotetext{a}{=S010202024704=U0179-0037591=2M00574957$-$7202361}
\tablenotetext{b}{=S010231095307=U0176-0047027=2M01003000$-$7220335}
\tablenotetext{c}{=S010202097167=0178-0047126=2M01012065$-$7211192}
\tablenotetext{d}{=S010202021404=U0179-0043444=2M01013695$-$7204148, 6$-$8\arcsec distant but considered as the counterpart by \citet{ha00,sa03}. }
\end{center}
\end{table}
\clearpage

\end{document}